\documentstyle[12pt]{article}
\begin{document}
\thispagestyle{empty}
\begin{raggedleft}
UR-1402\\
ER-40425-849\\
hep-th/9501081\\
Dec.\ 1994\\
\end{raggedleft}
$\phantom{x}$\vskip 0.618cm\par
{\huge \begin{center}COUPLING CHIRAL BOSONS TO GRAVITY
\footnote{This work is
supported by CNPq, Bras\'\i lia, Brasil}
\end{center}}\par
\begin{center}
$\phantom{X}$\\
{\Large N.R.F.Braga}\\[3ex]
{\em Instituto de F\'\i sica\\
Universidade Federal do Rio de Janeiro\\
21945, Rio de Janeiro, Brazil\\}
\end{center}\par
\begin{center}
$\phantom{X}$\\
{\Large Clovis Wotzasek\footnote{Permanent address:
Instituto de F\'\i sica,
Universidade Federal do Rio de Janeiro, Brasil}}\\[3ex]
{\em Department of Physics and Astronomy\\
University of Rochester\\
Rochester, NY 14627, USA}
\end{center}\par
\begin{abstract}

\noindent The chiral boson actions of Floreanini and Jackiw (FJ), and
of McClain,Wu and Yu (MWY) have been recently shown to be different
representations of the same chiral boson theory.  MWY displays
manifest covariance and also a (gauge) symmetry that is hidden in
the FJ side, which, on the other hand, displays the physical spectrum
in a simple manner.  We make use of the covariance of the MWY
representation for the chiral boson to couple it to background gravity
showing explicitly the equivalence with the previous results for the
FJ representation.

\end{abstract}
\vfill
\newpage

In recent years there has been a great deal of attention devoted to
the quantization of chiral scalar fields.  The main motivation
is that chiral bosons are the basic objects of  two of the
most interesting opened problems of present days theoretical physics.  They
appear in the construction of many string models \cite{string}
 where some symmetries are
manifest before chiral boson fermionization, but not after.
Technically, it is advantageous to keep chiral bosons, instead of their
fermionic counterparts, because it suffices to compute lower loop graphs on
the world-sheet.  Furthermore, in the description of quantum Hall effect
\cite{QHE},
chiral bosons play an important role since there they appear as the
edge-states of the Hall fluid, which are believed to be the only gapless
excitation of the sample.

To obtain a chiral boson one usually eliminates one half of the degrees of
freedom
from the scalar field by means of a chiral constraint $\partial_{\pm}
\phi \approx 0$.  The problem in following this route is that the chiral
constraint is second-class by Dirac's classification scheme
\cite{Dirac}.  Therefore one
is not allowed to gauge away
its associated Lagrange multiplier field that therefore acquires a
dynamical character.
Siegel \cite{Siegel}  proposed to covariantize the chiral constraint, i.e.,
to transform
it from second to first-class, by squaring it, or what is equivalent,
setting one chiral component of the energy-momentum tensor to zero
as a constraint.
Siegel's action has a reparametrization symmetry, at tree level, that
becomes anomalous at one-loop level (i.e., becomes second-class again after
quantization) \cite{anomaly1,anomaly2} due
to the existence of a central extension in the conformal algebra of the
energy-momentum tensor.  Later on, Hull \cite{hull} has shown how to cancel the
conformal anomaly of Siegel's model introducing auxiliary fields on the
zero-mode sector, the so called no-movers fields.
Nevertheless, to square a second-class constraint to
make it first-class results in a theory presenting (infinitely) reducible
constraints \cite{restucia}.  Independently, Floreanini and Jackiw
\cite{FJ}  proposed an action
where the chiral constraint appears from the equations of
motion and therefore does not involve any Lagrange multiplier field.  This
(first-order) chiral boson formulation introduces however some spurious
solutions to (second-order) field equations that need careful boundary
conditions adjustments to be eliminated.
Another drawback in FJ proposal is the lack of manifest Lorentz covariance,
which makes the coupling to gauge and gravitational fields difficult
\cite{Sonenschein}.

More recently, McClain, Wu and Yu \cite{MWY} have shown that an action
containing
infinite scalar fields, coupled by combinations of right and left chiral
constraints, carefully adjusted to be first-class, possess the spectrum of
a single chiral boson\footnote{The idea of using infinite auxiliary scalar
fields to covariantize second-class constraints has been introduced earlier
in the literature by Mikovic et al\cite{mik} in the context of the
relativistic super-particle.}.
  In fact, MWY and FJ are just two different
representations of the same chiral boson theory \cite{Clovis}
, each of which displaying
a different feature of the very same problem: in the MWY side not only
the Lorentz covariance is
manifest, but it also displays a symmetry that is hidden in the other
representation, while the FJ side, on the other hand, presents the
spectrum in a simpler manner.  Depending on one's interests, one
can pass from one representation to the other either transforming,
iteratively, the second-class constraint of the FJ model into
first-class, \`a la Faddeev-Shatashvili, with the
introduction of infinite Wess-Zumino fields, or
by resolving iteratively the MWY constraints by means of the
Faddeev-Jackiw technique for first-order constrained systems.

In this paper we shall make use of the manifest covariance of the MWY
representation to couple it to background gravity.  The results we
obtain are shown to be consistent with the ones previously obtained
for the FJ representation by Sonnenschein \cite{Sonenschein} and by
Bastianelli and van Nieuwenhuizen \cite{BN}, which are quickly
derived in an appendix, in a form slightly different then their
original formulation.

The MWY action is a representation of a  chiral boson in terms of
a sum over infinite scalar fields

\begin{equation}
\label{mwy}
S^{MWY}=\int d^2 x \sum_{k=0}^{\infty}(-)^k {1 \over 2} \partial_{\mu} \phi_k
\partial^{\mu}\phi_k
\end{equation}

\noindent chirally coupled to each other by a set of (infinite)
irreducible constraints $T_k^{(+)}$ or $T_k^{(-)}$ each one corresponding
to one of the chiralities

\begin{equation}
\label{mwy-constraint}
T_k^{(\pm)} = \Omega_k^{(\pm)}- \Omega_{k+1}^{(\mp)}
\end{equation}

\noindent with $\Omega^{(\pm)}_k$ being  right and left chiral constraints
$\Omega_k^{(\pm)}= \pi_k \pm \phi'$
that satisfy two uncoupled Kac-Moody algebra:

\begin{eqnarray}
\label{kac-moody}
\left\{\Omega_k^{(\pm)}(x),\Omega_m^{(\pm)}(y)\right\} & = & (\pm)2 \delta_{km}
\delta'(x-y)\nonumber\\
\left\{\Omega_k^{(+)}(x),\Omega_m^{(-)}(y)\right\} & = & 0
\end{eqnarray}

\noindent Here the curly brackets $\left\{A(x),B(y)\right\}$ represents the
Poisson bracket
of the fields $A(x)$ and $B(x)$.  One can verify that MWY constraints
$T_k^{(\pm)}(x)$ closes an (infinite) first-class Abelian algebra
under the Poisson bracket operation

\begin{equation}
\label{firstclass}
\left\{T_k^{(\pm)}(x),T_m^{(\pm)}(y)\right\}=0
\end{equation}

In view of the manifest covariance presented by the model, coupling to
gravity is straightforward, and reads

\begin{eqnarray}
\label{mwy-gravity}
S^{MWY}& = &{1\over 2}\int d^2 x \sqrt{-g}\sum_{k=0}^{\infty}(-)^k g^{\mu
\nu} \partial_{\mu}\phi_k\partial_{\nu}\phi_k\nonumber\\
& = & {1\over 2}\int d^2 x \sum_{k=0}^{\infty}
(-)^k\sqrt{-g} g^{00} \left\{\dot\phi^2_k +
2{g^{01} \over g^{00}}\dot\phi_k\phi'_k +{g^{11}\over g^{00}} \phi'^2_k
\right\}
\end{eqnarray}

\noindent where $g_{\mu \nu}$ is the background metric tensor and $g=\det
g_{\mu \nu}=g_{00}g_{11}-g_{01}^2$.  We adopt the usual notation where
$\dot\phi=\partial_0 \phi$ and $\phi' =\partial_1\phi$, and $x^0=\tau$ and
$x^1=\sigma$ are the two-dimensional world-sheet variables.
To obtain the gauged
Floreanini-Jackiw counterpart we have to reduce the MWY-constraints
(\ref{mwy-constraint}) as explained above.  In order to effect such a
reduction we make use of the Faddeev-Jackiw sympletic technique.  To this
end we rewrite the MWY action in its first-order form, by introducing
the momentum $\pi_k $ conjugate to $\phi_k$,

\begin{equation}
\label{1st-mwy-gravity}
S_{\pm}^{MWY} = \int d^2 x \sum_{k=0}^{\infty}\left\{\pi_k \dot\phi_k -
{\cal H}^{MWY}+ \lambda_k T_k^{(\pm)}\right\}
\end{equation}

\noindent where ${\cal H}^{MWY}$ is the canonical Hamiltonian density

\begin{equation}
\label{hamiltonian}
{\cal H}^{MWY}={1\over
2}\sum_{k=0}^{\infty} \left[{(-)^k \over{g^{00}\sqrt{-g}}}(\pi_k^2 +\phi_k'^2)
-{g^{01}\over g^{00}}\pi_k\phi_k'\right]
\end{equation}

\noindent Now we eliminate the momentum $\pi_k$, in an iterative fashion,
 making use of
the MWY-constraints.  Implementing the first constraint,
$\pi_0=\pi_1\mp\phi'_0\mp\phi'_1$ results, after its substitution, in the
following Lagrangian density

\begin{eqnarray}
\label{1st-iteration}
{\cal L}_{\pm}^{MWY}  &=& \mp\dot\phi_0\phi'_0 - {\cal G_{\pm}} \phi_0'^2 -
\phi'_1\left(\dot\phi_0 +{\cal G_{\pm}}\phi'_0\right)\nonumber\\
& &\mbox{} +  \pi_1\left[\dot\phi_0+\dot\phi_1+{\cal {G_{\pm}}}
(\phi'_0+\phi'_1)
\right]\nonumber\\
& &\mbox{} + \sum_{k=2}^{\infty}\left[\pi_k \dot\phi_k -{1\over
2}{(-)^k \over{g^{00}\sqrt{-g}}}(\pi_k^2 +\phi_k'^2) +{g^{01}\over
g^{00}}\pi_k\phi_k' \right]
\end{eqnarray}

\noindent where

\begin{equation}
\label{couplings}
{\cal G_{\pm}}={1 \over g^{00}}\left({1 \over
\sqrt{-g}}\pm g^{01}\right)
\end{equation}

\noindent We repeat this procedure for the constraint $\pi_1
=\pi_2\mp\phi'_1\mp\phi'_2$, in this way eliminating the momentum $\pi_1$, and
so on.  After all the remaining constraints have been implemented we find the
following effective action

\begin{equation}
\label{iterated1}
{\cal L}_{\pm}^{MWY}=\sum_{k=0}^{\infty}\left[\mp\dot\phi_k\phi'_k-
{\cal G_{\pm}}
\phi_k'^2 -2\left(\dot\phi_k +{\cal G_{\pm}}\phi'_k\right)
\sum_{m=k+1}^{\infty}\phi'_m\right]
\end{equation}

\noindent It is a simple algebraic manipulation to rewrite this action as
\begin{equation}
\label{iterated2}
{\cal L}_{\pm}^{MWY}=\sum_{k=0}^{\infty}\left[\mp\dot\phi_k\phi'_k-
{\cal G_{\pm}}
\phi_k'^2 -2\phi'_k\sum_{m=1}^{k-1} \left(\dot\phi_m +
{\cal G_{\pm}}\phi'_m\right)\right]
\end{equation}

\noindent which shows that all the MWY scalar fields have decoupled from
each other.  To make this point clearer, we rewrite the action
(\ref{iterated2}) as a double series function

\begin{equation}
\label{doubleseries}
{\cal S}_{\pm}^{MWY}= \sum_{k=0}^{\infty}\sum_{m=0}^{\infty}\int d^2x
\left(\mp\dot\phi_k\phi'_m - {\cal
G_{\pm}}\phi'_k\phi'_m \right)
\end{equation}

\noindent and introduce a new (collective) variable as

\begin{equation}
\label{collective}
\Phi = \sum_{k=0}^{\infty} \phi_k
\end{equation}

\noindent The MWY action for the collective field $\Phi$ assumes the form
of a Floreanini-Jackiw action coupled to the gravitational field
by the factor  ${\cal G_{\pm}}$

\begin{equation}
\label{florjack}
{\cal S}_{\pm}^{MWY} = \int d^2x\left(\mp\dot\Phi\Phi' -
{\cal G_{\pm}}\Phi'^2\right)
\end{equation}

An interesting feature of the interacting action
(\ref{florjack}) is the reparametrization symmetry
that the coupling  ${\cal G_{\pm}}$ induces on the system, that reads

\begin{eqnarray}
\label{symmetry}
\delta_{\epsilon} \Phi &=& \epsilon \Phi'\nonumber\\
\delta_{\epsilon} {\cal G_{\pm}} &=& (\dot\epsilon + \epsilon {\cal G_{\pm}}' -
\epsilon' {\cal G_{\pm}})
\end{eqnarray}

It is also interesting to note that for each of the chiral fields,
$\phi_+$ or $\phi_-$, there will exist a class of metrics
for which they propagate as in flat space.
This will be the case whenever ${\cal G}_+ = 1$ or ${\cal G}_- = 1 $,
respectively.
These two conditions corresponds, as can be seen from (\ref{couplings}),
to respectively, $g^{00} + g^{11} \pm 2 g^{01}=0$, or in terms of
light cone coordinates, to:

\begin{eqnarray}
\label{conditions}
g^{--} &=& 0\nonumber\\
g^{++} &=& 0.
\end{eqnarray}

\noindent A symmetric two dimensional metric can be written in terms of
the light cone components in the general form

\begin{equation}
\label{metric}
g^{\mu\nu}={1\over 2} \left(
\begin{array}{cc}
g^{++} + g^{--}+2 g^{+-} & g^{++} - g^{--} \\
g^{++} - g^{--} & g^{++} + g^{--} -2 g^{+-}
\end{array}
\right)
\end{equation}

\noindent Therefore, in metrics of the form

\begin{equation}
\label{metric+}
g^{\mu\nu} ={1 \over 2} \left(
\begin{array}{cc}
g^{++} + g^{+-} & g^{++} \\
g^{++} & g^{++} - g^{+-}
\end{array}
\right)
\end{equation}

\noindent which gives ${\cal G_+}=1$, the chiral field $\phi_+$ propagates
as in flat space.  Similarly, for metrics whose general form reads

\begin{equation}
\label{metric-}
g^{\mu\nu} = \left(
\begin{array}{cc}
g^{--} + g^{+-} & -g^{--} \\
-g^{--} & g^{--} - g^{+-}
\end{array}
\right)
\end{equation}

\noindent implying ${\cal G_-}=1$, the chiral field $\phi_-$ remain
uncoupled.  Stated differently, chiral bosons, when immersed in a
curved background, select to couple to a special combination of the
metric elements ${\cal G_{\pm}}$, in a similar way as it happens when
coupling to gauge fields.  Under conditions (\ref{conditions}) the
metrics (\ref{metric+}) and (\ref{metric-}) become, in a sense,
``chiral metrics''.  Consequently when immersed in a chiral
background with chirality opposite to its own, chiral boson just do
not experiment the curvature. It should be noted that, contrarily to
the case of gauge fields, where the coupling is additive, when
conditions (\ref{conditions})are not satisfied each of the chiralities
of the field $\phi$ couples to the corresponding ${\cal G_{\pm}}$ that
depends on the whole metric and not on (\ref{metric+}) or (\ref{metric-})

  Conditions (\ref{conditions}) correspond to a sort of selfduality
and anti-selduality over the metric tensor, justifying us to call
them as chiral metrics.  Indeed, if we define the ``dual metric''
with respect to one index, as

\begin{equation}
\label{dual}
\mbox{}^{*}g^{\mu\nu}=\epsilon^{\mu\lambda}g_{\lambda}^{\nu}
\end{equation}

\noindent with $\epsilon^{10}=-\epsilon^{01}=1$, and take the trace, then
conditions (\ref{conditions}) will read

\begin{equation}
\label{trace}
Tr\,\left(g^{\mu\nu}\right)=\pm Tr\left(\mbox{}^{\ast} g^{\mu\nu}\right)
\end{equation}

\noindent as claimed

Concluding, we have shown explicitly that the MWY representation for
the chiral boson can be coupled to an external background metric
in a standard way, making use of the manifest covariance.
The incorporation of the series of infinite chiral constraints in the model
was shown to lead to the same result as the one previously obtained for
the FJ model.
We have also pointed out the classes of metrics for which one of the
chiralities propagate as in free space.

\appendix
\renewcommand{\theequation}{\thesection.\arabic{equation}}
\section{Appendix}
\setcounter{equation}{0}

In reference \cite{BN} it was shown that a version of the Floreanini Jackiw
action, coupled to gravity can be obtained beginning with a scalar field
coupled covariantly to gravity then writing it as a first order action
and imposing a (non covariant) constraint that selects one of the
(chiral) solutions of the classical equation of motion. Here we will
derive the same result using the Mandelstan
\cite{Mandelstan}
decomposition for a scalar field into chiral bosons.  Let us begin,
as \cite{BN}, with a scalar
field coupled covariantly to gravity:

\begin{equation}
\label{scalar}
{\cal L} = {1 \over 2} \sqrt{-g} g^{\mu\nu} \partial_{\mu} \phi
\partial_{\nu}\phi
\end{equation}

\noindent The introduction of the auxiliary variable $p$ makes it possible to
write this Lagrangian density in a first order form

\begin{equation}
\label{scalarfirst}
{\cal L} = {\sqrt{-g}\over 2} \left[ -g^{00} p^2
+ 2 g^{00} p \dot\phi
+ 2g^{01} \dot\phi \phi^{\prime} +  g^{11} \phi^{\prime}\phi^{\prime}
\right]
\end{equation}

\noindent Decomposing the scalar field in its (Mandelstan) components:

\begin{equation}
\label{mandelstan}
\phi = \phi_+ + \phi_-
\end{equation}

\noindent one is able to associate each of this fields with one
of the (chiral) solutions of the classical equation of motion by
imposing\cite{remark}:

\begin{equation}
\label{mandelstan2}
p ={g^{01}\over g^{00}} ( \phi_+^{\prime} + \phi_-^{\prime} )
+ {1\over \sqrt{-g} g^{00}} ( \phi_-^{\prime} - \phi_+^{\prime})
\end{equation}

\noindent Inserting (\ref{mandelstan}) and (\ref{mandelstan2}) in
(\ref{scalarfirst}) we get

\begin{eqnarray}
\label{scalarmandelstan}
{\cal L}  & = & -\dot\phi_+ \phi_+^{\prime} -
{\cal G_+} \phi_+^{\prime} \phi_+^{\prime} \nonumber\\
& &\mbox{} + \dot\phi_- \phi_-^{\prime} -
{\cal G_-} \phi_-^{\prime} \phi_-^{\prime}
\end{eqnarray}

\noindent showing explicitly the decomposition of the scalar field in
chiral components, each of them coupling with the metric exactly
as in \cite{BN}.

\end{document}